\newcommand{\sla}{\not\!\!}
\documentstyle[12pt]{article}
\begin{document}
\begin{titlepage}
\begin{flushright}
NSF-ITP-93-68\\
TPI-MINN-93/33-T\\
UMN-TH-1208/93\\
July 1993\\
\end{flushright}
\vspace{0.4in}
\begin{center}
{\Large \bf Differential Distributions in Semileptonic Decays of the
Heavy Flavors in QCD}
\end{center}
\vspace{0.2in}
\begin{center}
{\bf B. Blok} \footnote{e-mail address: BLOK@SBITP.BITNET} \\
Institute for Theoretical Physics\\
University of California, Santa Barbara, CA 93106\\
and\\
{\bf L. Koyrakh \footnote{e-mail address: LEVKOYRAKH@PHYSICS.SPA.UMN.EDU},
 M. Shifman \footnote{e-mail address: SHIFMAN@UMNACVX}
 and  A.I. Vainshtein \footnote{and Budker Institute of Nuclear Physics,
Novosibirsk 630090, Russia} \footnote{e-mail address: VAINSHTE@UMNACVX}} \\
Theoretical Physics Institute\\
University of Minnesota, Minneapolis, MN 55455\\
\end{center}
\begin{center}
\vspace{0.2in}
{\large \bf Abstract}
\end{center}
{A generalization of the operator product expansion is
used to find the differential distributions in  the  inclusive semileptonic
weak decays of heavy flavors in  QCD. In particular, the double
distribution in electron energy and invariant mass of the lepton pair is
calculated. We are able to calculate the distributions in essentially
model-independent way as a series in $m_Q^{-1}$  where $m_Q$ is the
heavy quark mass. All effects up to $m_Q^{-2}$ are included.
}
\vfill
\end{titlepage}
\section{Introduction}
Differential distributions in semileptonic decays of heavy flavors
are used for measurements of the CKM matrix elements, key phenomenological
parameters of the standard model. To extract the CKM matrix elements from
data one needs to disentangle the effects of strong interactions at large
distances from the quark-lepton lagrangian known at short distances.

Up to now essentially two  approaches are applied to describe
nonperturbative
strong interaction effects in the inclusive weak decays: the
naive parton model amended to include the motion of the heavy quark
inside the decaying meson \cite{Alta1}; and the 'exclusive variant' based on
summation of different channels, one by one \cite{Isgu1}.
Both approaches are admittedly model-dependent, neither their accuracy nor
the connection
to the fundamental parameters of QCD are clear {\em a priori}.
Each of them needs an input from constituent
quark model to parametrize nonperturbative effects. The latter
play an especially important role in the form of the spectra near the
endpoints.

The need for the model-independent QCD-based predictions is apparent.\\
Considerable progress achieved recently in the theory of preasymptotic
effects (proportional to powers of $1/m_Q$ where $m_Q$ is the heavy quark
mass) allows one to make these
predictions.

The theoretical construction presented in this paper is, in a sense, a
generalization and combination of the formalisms which are
used in deep inelastic scattering and total cross section of $e^+e^-$
annihilation. The
expansion parameter in deep inelastic scattering is $Q^{-1}$ where $Q$ is
the momentum transfer. In the problem at hand the expansion parameter is
$m_Q^{-1}$ or, more exactly, the inverse energy released in the final
hadronic state (in the rest frame of the decaying quark).

In the classical problems of  this type, like $e^+e^-$ -- annihilation,
there are two alternative ways to get predictions. The first approach
having a solid theoretical justification in terms of OPE \cite{Wils1}
is based on calculations in the euclidean domain where one can apply OPE.
The contact with the observable quantities is made through the dispersion
relations and in this way predictions for certain integrals are obtained.
In the second approach we perform the calculations directly in Minkowski
domain. Although formally this calculation refers to large distances from
the first approach we know that in specific integrals large distance
contributions drop out. Therefore the results obtained in this way, although
not valid literally, should be understood in the sense of duality: being
smeared over some duality interval the theoretical prediction should coincide
with the smeared experimental curve. The inclusive weak decays will be
treated within the second approach. The averaging mainly refers to
the invariant mass of the inclusive hadronic state produced in the decay
considered.

If the invariant mass of the final hadronic state is large this is not a
constraint at all since the theory 'itself' takes care of the averaging
required by duality. In the opposite limit, near a spectral endpoint,
the smearing is not provided for free. We will discuss the issue in some
detail below.

Although we explicitly work in the Minkowski kinematics we always keep in mind
the relationship to the euclidean domain and the corresponding
operator product expansion. The first analysis of this type has
been outlined in \cite{Volo1} for inclusive heavy flavor decay
rates. A general analysis of the semileptonic inclusive spectra along this
line is presented in ref. \cite{Geor2}.
In that work it was observed, in particular, that the leading
operator and those appearing at the next-to-leading order have a gap in
dimensions of two units, and, consequently, the ${\cal O}(m_Q^{-1})$ term
should be absent in certain quantities.  The  analysis presented
in \cite{Geor2} was not backed up, however, by concrete calculations of the
preasymptotic effects.
Recently this formalism has been systematically developed and
applied to the non-leptonic decays of heavy flavors
\cite{Bigi1,Blok1} and the charged-lepton energy spectrum in the semileptonic
decays \cite{Bigi4} (see also \cite{Bigi3}).
The present work is a natural continuation of ref. \cite{Bigi4}.

We generalize the results of ref.\cite{Bigi4} to find the complete inclusive
distributions in the semileptonic decays. The leptonic variables --- $E_e,
q^2$ and $q_0$, where $E_e$ is the charged lepton energy and $q$ is the
momentum of the lepton pair \footnote{The charged lepton produced will be
generically called 'electron' hereafter.}  --- are kept fixed which
automatically fixes the invariant mass of the inclusive hadronic state.
Integrating over $q_0$ we obtain the double spectral distribution in $E_e$
and $q^2$.

At the first stage we construct the transition
{\em operator} $T(Q\rightarrow X\rightarrow Q)$ describing the
forward scattering amplitude of the heavy quark $Q$ on a weak
current. Our focus is the influence of the 'soft' modes (background fields)
on the transition operator $T_{\mu\nu}$ which is expressed  as an infinite
series in the local operators built from the gluon and/or light-quark fields
bilinear in $Q\bar Q$.

The local operators are ordered according to their dimensions;
the coefficient functions contain the corresponding powers of $1/m_Q$
(or $1/E$, the energy release). At sufficiently
large $m_Q$ or $E$ the operators with the lowest dimensions dominate,
and the infinite series can be truncated.

At the next stage the matrix elements of the relevant operators over the
initial heavy hadron $H_Q$ must be
evaluated. Unfortunately, in the present-day QCD the matrix elements over
the hadronic states are not theoretically calculable. In some instances
they can be related, through heavy quark symmetries, to measurable
quantities \cite{Geor1}, \cite{Isgu1};
 in other cases they have
to be parametrized.  These parameters play the role analogous to the gluon
condensate \cite{Shif1}. As a matter of fact, at the level of the leading
preasymptotic corrections only two operators are relevant. The matrix element
of the first one can be related to the mass splittings of the vector and
pseudoscalar heavy mesons. The matrix element of the second one has the
meaning of the average square of the spatial momentum of the heavy quark
$Q$ in $H_Q$ and the state must be treated as a parameter.

Finally, the observed decay rates and spectra are obtained by taking the
discontinuity of the hadronic tensor $<H_Q|T_{\mu\nu}|H_Q>$ and convoluting
the result with the lepton currents and appropriate
kinematic factors.

In this paper we consider the differential distributions in
the semileptonic decays at the level of $O(m_Q^{-2})$.
The differential distributions are measured experimentally in the $B$
meson decays and will be used for more precise determination of $V_{ub}$,
for example. This was a primary motivation for our  investigation.
We would like to make it as close to the fundamental QCD as possible.

The organization of the paper is as follows.
In Sect. 2  we describe the kinematics and in Sect. 3 we present the
operator product expansion. In Sect. 4 we derive the
differential distributions. Sect. 5 is devoted to the
 analysis of our distributions  and  limitations on the their use.
Our results are summarized in  Sect. 6. Appendix contains expressions for
hadronic invariant functions.

\section{Kinematical analysis}
We will consider the inclusive weak decays of the mesons (or baryons)
with the open heavy flavor into the lepton pair plus (inclusive) hadronic
state $H_Q(p_H) \rightarrow l(p_l)+\bar\nu(p_\nu)+hadrons$.
 Our final goal is to calculate the differential decay rate
\begin{equation}
d^3\Gamma\over dE_edq^2dq_0
\label{eq:d3G}
\end{equation}
where $E_e$ is the energy of the emitted electron
and $q^\mu=p_l^\mu+p_\nu^\mu$ is the 4-momentum of the lepton pair.
In order to find the differential distributions we need to know the amplitude
of the process, which is given by the expression
\begin{equation}
{\cal M} =V_{qQ}\,{G_F\over \sqrt{2}}\;\bar e \,\Gamma_\nu\,\nu\,
<X|j_\nu|H_Q>
\label{eq:amplitude}
\end{equation}
Here $V_{qQ}$ is the corresponding Cabibbo-Kobayashi-Maskawa matrix element,
$j_\mu=\bar q\Gamma_\mu Q$ is the electroweak currents,
$\Gamma_\mu=\gamma_\mu(1+\gamma_5)$.  (Although our theory is general we will
keep in mind the $b\rightarrow c$ and $b\rightarrow u$ decays, so that
$Q=b$ and $q=c$ or $u$).
The differential distributions we are interested in are given by the
modulus squared of the amplitude (\ref{eq:amplitude}) summed over the
 final hadronic states.

The modulus squared of the amplitude summed over the final hadronic states
can be written as
\begin{equation}
|{\cal M}|^2 =|V_{qQ}|^2 G_F^2 M_{H_Q}\, l^{\mu\nu}\,W_{\mu\nu}
\label{eq:amp2}
\end{equation}
where $M_{H_Q}$ is the mass of hadron
$H_Q$, $W_{\mu\nu}$ is the hadronic tensor
\begin{equation}
W_{\mu\nu} = (2\pi)^4\,\delta^4(p_{H_Q} - q - p_X)\,
 \sum_X {1\over 2 M_{H_Q}}<H_Q|j_\mu^{\dag}(0)|X><X|j_\nu(0)|H_Q>,
\label{eq:Wtens}
\end{equation}
and $l^{\mu\nu}$ is the lepton tensor
\begin{equation}
l^{\mu\nu}=8\,[\,(p_{e})^\mu (p_{\nu})^\nu + (p_{e})^\nu (p_{\nu})^\mu -
        g^{\mu\nu}(p_e\cdot p_\nu ) + i\epsilon^{\mu\nu \alpha\beta }
(p_e)^\alpha(p_\nu )^\beta].
\label{eq:lmunu}
\end{equation}

Let us introduce the hadronic structure functions $w_i$ and parametrize
the hadronic tensor in the following way:
\begin{eqnarray}
W_{\mu \nu} =
     -w_1\,g_{\mu \nu}
     +w_2\,v_\mu\,v_\nu
     -i\, w_3\,\epsilon_{\mu\nu\alpha\beta}\,v_\alpha q_\beta + \nonumber\\
     w_4\,q_\mu\,q_\nu + w_5\,(q_\nu\,v_\mu+q_\mu\,v_\nu).
\label{eq:W2w}
\end{eqnarray}
  Here $q_\mu=(p_e+p_\nu)_\mu$ is the 4 - momentum of the
lepton pair, $v_\mu=(p_{H_Q})_\mu / M_{H_Q}$ is the 4-velocity of the
initial {\em hadron}.
Note that we have omitted the structure $q_\mu v_\nu-q_\nu
v_\mu$ which can not appear because of the $T$--invariance.  The structure
 functions $w_i$ depend on two invariant variables, $q\cdot v$ and
$q^2$. In the rest frame of $H_Q$ which will be used throughout the paper
$q\cdot v = q_0 $, so $w_i=w_i(q_0,q^2)$.
The convolution of $W_{\mu\nu}$ with the lepton tensor (\ref{eq:lmunu}) is
given by the expression:
\begin{equation}
W_{\mu\nu}\,l^{\mu\nu} = 4\,\{
2\,q^2\,w_1 + [\,4\,E_e\,(\,q_0 - E_e\,) - q^2\,]\,w_2 + 2\,q^2\,(\,2\,E_e -
  q_0)\,w_3 \,\}.
\label{eq:wl} \end{equation}
We see that only three
invariant functions are relevant for the processes we are considering
in this paper.
At this step we encounter the third variable, the electron energy
$E_e=p_e\cdot p_{H_Q}/M_{H_Q}$, entering through the leptonic tensor.

Finally the formulae for the differential width takes the form
\begin{equation}
{d^3\Gamma \over dE_e\,dq^2dq_0} = |V_{qQ}|^2 {G_F^2\over 32\,\pi^2}\,
[2 q^2 w_1+ [4\,E_e(q_0-E_e)-q^2]\,w_2+2\,q^2(2\,E_e-q_0)\,w_3\,]
\label{eq:d3}
\end{equation}

This expression concludes the kinematical analysis. Our task is,
of course, the calculation of the invariant functions $w_i(q_0,q^2)$.
We will proceed to this calculation in the next section.

\section{Operator product expansion}

In this section we will discuss the derivation of the tensor $W_{\mu\nu}$.
The Operator product expansion is similar to that in the deep inelastic
scattering. It is convenient to introduce the hadronic tensor $h_{\mu\nu}$
(forward scattering amplitude)
\begin{equation}
h_{\mu\nu}=i\int
d^4xe^{-iqx}{1 \over 2\,M_{H_Q}}<H_Q\vert T\{ j_\mu^+(x) j_\nu(0)\} \vert H_Q>.
\label{eq:hmunu}
\end{equation}
The absorptive part of this tensor reduces to $W_{\mu\nu}$ discussed above
\begin{equation}
W_{\mu\nu} = (1/i)\,{\rm disc}(h_{\mu\nu}).
\label{eq:Wh}
\end{equation}
Here  ${\rm disc}(h_{\mu\nu})$ is the discontinuity of the forward
scattering amplitude  $h_{\mu\nu}$ on the physical cut in the complex
plane of the variable $q_0$. Of course, $h_{\mu\nu}$ can be expanded
into the same set of structures as $W_{\mu\nu}$ (see. eq. (\ref{eq:W2w}))
\begin{eqnarray}
h_{\mu \nu} =
     -h_1\,g_{\mu \nu}
     +h_2\,v_\mu\,v_\nu
     -i\, h_3\,\epsilon_{\mu\nu\alpha\beta}\,v_\alpha q_\beta +
     h_4\,q_\mu\,q_\nu + h_5\,(q_\nu\,v_\mu+q_\mu\,v_\nu),
\label{eq:h2h}
\end{eqnarray}
and the relation (\ref{eq:Wh}) implies that
\begin{equation}
w_i = 2\,{\rm Im}(h_i).
\label{eq:h2w}
\end{equation}

Let us remind that $h_{\mu\nu}$ is the matrix element of the transition
operator $T_{\mu\nu}$
\begin{equation}
h_{\mu\nu} = {1 \over 2\,M_{H_Q}}<H_Q| T_{\mu\nu}|H_Q>,
\end{equation}
\begin{equation}
T_{\mu\nu} = i\,\int d^4x e^{-iqx} \, T\{ j_\mu^+(x) j_\nu(0)\},
\label{eq:Tmunu}
\end{equation}
so below we will construct OPE for the product of currents in equation
(\ref{eq:Tmunu}). Having in mind the relationship to euclidean analysis
discussed above we will treat our expansion in the same way as a
normal euclidean OPE.
In the asymptotic limit $m_Q \rightarrow \infty$ the hadronic tensor
$h_{\mu\nu}$ is given by the tree graph of Fig. 1. This graph defines the
matrix element of the transition operator $T_{\mu\nu}$ over the heavy
quark state,
\begin{equation}
<Q|T_{\mu\nu}|Q> = -\bar u_Q \Gamma_\mu \frac{1}{\sla P-\sla\, q\
 - m_q}\Gamma_\nu\,u_Q
\label{eq:QTQ}
\end{equation}
This expression represents nothing else but the free quark decay.
In the asymptotic regime $m_Q \rightarrow \infty$ the
interaction of the heavy quark with the gluon/light quark medium, as well
as its intrinsic notion inside the hadron can be neglected. Then
\begin{equation}
P_\mu=P_{0\,\mu} \equiv m_Q\,v_\mu
\label{eq:PP0}
\end{equation}
where $v_\mu$ is the $H_Q$ 4-velocity.

Equation (\ref{eq:QTQ}) allows one to immediately write down the operator
form in the approximation at hand (only operators bilinear in $Q$, $\bar Q$
are considered, see a discussion of other operators at the end of the section):
\begin{eqnarray}
T_{\mu\nu} & = & -\bar Q \Gamma_\mu \frac{1}{\sla k - m_q}
\Gamma_\nu\,Q = \\
          &   & - { 2 \over (k^2-m_q^2)}[g_{\alpha\mu}k_\nu +
g_{\alpha\nu}k_\mu - g_{\mu\nu}k_\alpha -
i \epsilon_{\mu\nu\alpha\beta}k_\beta]\,
\bar Q \gamma^\alpha (1+\gamma_5)Q \nonumber
\label{eq:TOPE}
\end{eqnarray}
where $k=P_0 - q$.

As we see, the two operators $\bar Q\gamma_\alpha\bar Q$ and
$\bar Q\gamma_\alpha\gamma_5\bar Q$ showed up in the operator expansion
at the level considered. Note, that the $\bar Q\gamma_\alpha\gamma_5\bar Q$
term vanishes after averaging over the unpolarized hadronic states.

In this paper the perturbative corrections in $\alpha_s$ are not
touched upon at all. As for nonperturbative corrections they appear
due to interactions with the soft medium  of the light cloud in $H_Q$.
By taking these interactions into account we isolate two types of effects.
First, the fast quark $q$ produced does not propagate as a free one, but
interacts with the background fields; these corrections will be
included explicitly. Second, the heavy quark $Q$ also does not live in
the empty space; it is surrounded by the light cloud. In particular, due
to this fact the heavy quark momentum does not coincide with $m_Q v_\mu$.
This large distance effect will not be calculated explicitly, but implicitly
it will be reflected in the $H_Q$ matrix elements of the operators
in $T_{\mu\nu}$. This is in full analogy with what people usually do in the
deep inelastic scattering. The influence of the background field on the
transition operator is summarized by the following expression
\begin{equation}
T_{\mu\nu} = -\int dx e^{-iqx} \bar Q(x) \Gamma_\mu S_q(x,0)\,\Gamma_\nu\,Q(0)
\label{eq:TSq}
\end{equation}
where $S_q(x,0)$ is the propagator of the quark $q$ in an external gluon
field $A^a_\mu$. It is convenient to use the Schwinger technique of treating
the motion in an external field (for a review of QCD adaptation see ,
e.g. ref. \cite{Novi1}). Within that formalism the propagator $S_q$ is
presented by the following expression
\begin{equation}
S_q(x,0)=(x|{1 \over \sla {\cal P} -m_q} |0)
\label{eq:Sq}
\end{equation}
Here $\sla {\cal P} = \gamma^\mu(p_\mu+A_\mu(X))$,  $A_\mu=g\,A_\mu^a\,T^a$
is a gluon field in the matrix representation. Furthermore, the operator
of coordinate $X_\mu$ and
momentum $p_\mu$ are introduced, (thus the field $A_\mu(X)$ becomes an
operator function of $X_\mu$), with the commutation relations
\begin{equation}
[p_\mu,X_\nu] =i\,g_{\mu\nu}, \quad [X_\mu,X_\nu] = 0,\quad [p_\mu,p_\nu] = 0.
\label{eq:px}
\end{equation}
The states $|x)$ are the eigenstates of the
operator $X_\mu$, $X_\mu|x)=x_\mu|x)$.

Combining equations (\ref{eq:Sq}) and (\ref{eq:TSq}), we arrive at
\begin{equation}
T_{\mu\nu} = -\int dx e^{-iqx}
(x|\bar Q(x) \Gamma_\mu {1 \over \sla {\cal P} -m_q }\,\Gamma_\nu\,Q(X)|0).
\label{eq:Tint}
\end{equation}

As we have discussed above the operator ${\cal P}_\mu$ contains a large
mechanical part $(P_0)_\mu=m_Q v_\mu$; the deviation from $P_0$ will be
separated explicitly
\begin{equation}
{\cal P}_\mu= (P_0)_\mu+\pi_\mu
\end{equation}
and we will expand in $\pi_\mu$. In this paper we will limit ourselves to the
terms up to ${\cal O}(\pi^2)$ corresponding to $1/m_Q^2$ corrections.
The master formulae to perform the expansion is
\begin{equation}
T_{\mu\nu} = -\int dx
(x|\bar Q(X) \Gamma_\mu {1\over\sla P_0-\sla q-m_q+\sla \pi}\,
\Gamma_\nu\,Q(X)|0).
\label{eq:master}
\end{equation}

There is a subtle point in the description of the formalism given above.
Technically in the computation the $A_\mu(x)$ is assumed to be a c-number
background field while in the final expression for local operators it should
be understood as a second quantized operator. Since we are not considering
any loop corrections this substitution is justified.

Let us now discuss the set of the operators relevant to the order
${\cal O}(m_Q^{-2})$. Without loss of generality we can work in the
rest frame of the hadron $H_Q$, i.e. $v_\mu=(1,0,0,0)$. Only those
operators will be retained which produce non-vanishing results after
being averaged over $H_Q$. The leading operator, as it was discussed above,
is
\begin{equation}
\bar Q \gamma_0 Q,
\label{eq:QgamQ}
\end{equation}
it's matrix element is fixed by the vector current conservation to be
\begin{equation}
{1\over 2\,M_{H_Q}}<H_Q|\bar Q \gamma_0 Q|H_Q>=1,
\label{eq:HQgamQH}
\end{equation}
Equation (\ref{eq:HQgamQH}) is given in relativistic normalization we are
using throughout this paper. In the
non-relativistic normalization there is no need in the factor
$1/2M_{H_Q}$ in the l.h.s.

As it has
been noted in the ref. \cite{Geor2} there are no operators of dimension 4
in the problem at hand. The set includes two operators of dimension 5:
\begin{equation}
{\cal O}_G= {i\over2}\bar Q \sigma^{\alpha\beta}G_{\alpha\beta} Q,
\label{eq:OG}
\end{equation}
\begin{equation}
{\cal O}_\pi  = -\bar Q\,\vec D \,^2 Q = \bar Q \,\vec \pi\,^2 Q,
\label{eq:Opi}
\end{equation}
where $\sigma^{\alpha\beta}={1\over 2}\, (\gamma^\alpha \gamma^\beta
-\gamma^\beta \gamma^\alpha)$, and $G_{\alpha\beta}=g\,G_{\alpha\beta}^q T^a$
is the gluon field strength tensor. The classification above takes
into account the fact that the quark field $Q$
satisfies the equation of motion. In
particular, it stems that the operator $\bar Q Q$ is not independent but is
reducible to three operators (\ref{eq:QgamQ}),  (\ref{eq:OG}) and
(\ref{eq:Opi}):
\begin{equation}
\bar Q\,Q = \bar Q\,\gamma_0 Q -
{1\over 2\,m_Q^2}\,\bar Q \,\vec \pi\,^2 Q +
{i\over 4\,m_Q^2}\,\bar Q \sigma^{\alpha\beta}G_{\alpha\beta} Q +
{\cal O}(m_Q^{-3}).
\label{eq:QQ}
\end{equation}
To get eq. (\ref{eq:QQ}) we observe that the lower component of $Q$
is related to the upper one in the following way
\begin{equation}
\frac{1-\gamma_0}{2}\,Q = \frac{1}{2m_Q} \vec \pi \vec \sigma
\frac{1+\gamma_0}{2}\,Q + {\cal O}(m_Q^{-2}),
\end{equation}
and the difference between $\bar Q Q$ and $\bar Q \gamma_0 Q$ is due to
the product of the lower components.
(Here and below we will stick to the $H_Q$ rest frame.)

A few other useful relations which can be obtained in the same manner and
are valid at the level ${\cal O}(m_Q^{-2})$ are:
\begin{equation}
\bar Q \vec\gamma \vec\pi  Q = \frac{1}{m_Q} \bar Q ({\vec\pi}^2
- {i \over 2} \sigma G) Q + {\cal O}(m_Q^{-2}),
\label{eq:QgampiQ}
\end{equation}
\begin{equation}
\bar Q \vec\gamma \vec\pi\gamma_0  Q = {\cal O}(m_Q^{-2}),
\label{eq:Qgampigam0Q}
\end{equation}
\begin{equation}
\bar Q \pi_0 Q = \frac{1}{2m_Q} \bar Q ({\vec\pi}^2
- {i \over 2} \sigma G) Q + {\cal O}(m_Q^{-2}).
\label{eq:Qpi0Q}
\end{equation}

A few comments are in order here concerning the actual technique of
constructing the OPE.
Since we work in the $H_Q$ rest frame it is
convenient to compute different components of $T_{\mu\nu}$ separately,
$T_{00},\,\,\, T_{0i},\,\,\, T_{i0}$ and $T_{ij}$.
The calculation itself is a straightforward although rather tedious
procedure of expanding  the denominator in eq. (\ref{eq:master})
in $\sla\,\pi$
using the properties of the $\gamma$ matrices, the commutation relation
\begin{equation}
[\pi_\mu,\pi_\nu]\,=\,i\,G_{\mu\nu}
\label{eq:pimupinu}
\end{equation}
and equations (\ref{eq:QgampiQ}) - (\ref{eq:Qpi0Q}).

Notice that we must keep the terms of the first order
in $\pi_0$ and of the second
order in $\vec \pi$, since
\begin{equation}
\pi_0\,Q={(\vec\sigma\vec\pi)^2\over 2 m_Q}Q + {\cal O}(m_Q^{-2}).
\label{eq:pi0Q}
\end{equation}
Next, observe that the Green function in the background field can be
written as follows:
\begin{equation}
\frac{1}{\sla{\cal P } -\sla \,q-m_q}=
(\sla{\cal P } -\sla\, q +m_q)\frac{1}{({\cal P }-q)^2+(i/2)\sigma G
-m_q^2}\equiv (\sla{\cal P } -\sla\, q +m_q)\frac{1}{\Pi}.
\end{equation}
To transpose $1/ \Pi$ with $\Gamma_\nu$ it is convenient to use the
identity
$$
\frac{1}{\Pi}\Gamma_\nu =\Gamma_\nu\frac{1}{\Pi} +
\frac{1}{\Pi}\,[\Gamma_\nu,\Pi]\,\frac{1}{\Pi}
$$
\begin{equation}
= \Gamma_\nu\frac{1}{\Pi}+\frac{1}{\Pi}[\Gamma_\nu ,{i\over 2}\sigma G]
\frac{1}{\Pi}.
\label{eq:PI}
\end{equation}
Acting on $Q$ and using the equations of motion we can now substitute
$1/\Pi$ in both terms on the right-hand side by
\begin{equation}
\frac{1}{m_Q^2-m_q^2-2{\cal P}q +q^2},
\end{equation}
provided that we limit ourselves to terms up to ${\cal O}(m_Q^{-2})$.
The second term in (\ref{eq:PI}) can be simplified even further
since here we can
additionally neglect $\pi$ in ${\cal P}= P_0 +\pi$.
\par We split the calculation into three parts:
Vector$\times$Vector, Axial$\times$Axial and Vector$\times$Axial in
correspondence with the structure of $\Gamma_\mu$ as a sum of vector
and axial vector, $\Gamma_\mu=\gamma_\mu + \gamma_{\mu}\gamma_5$.
The full hadronic tensor $h_{\mu\nu}$ is given then by the following
expression:
\begin{equation}
h_{\mu\nu}=
h^{VV}_{\mu\nu}+h^{AA}_{\mu\nu}+h^{AV}_{\mu\nu}+h^{VA}_{\mu\nu} =
{1\over 2\,M_{H_Q}}
<H_Q|T^{VV}_{\mu\nu}+T^{AA}_{\mu\nu}+T^{AV}_{\mu\nu}+T^{VA}_{\mu\nu}|H_Q>\,.
\label{eq:full}
\end{equation}
The complete expressions for the hadronic invariant functions are
given in the Appendix. In the order ${\cal O}(m_Q^{-2})$ they are defined
by matrix elements of operators ${\cal O}_G$, ${\cal O}_\pi$ given by
eqs.(\ref{eq:OG}), (\ref{eq:Opi}):
\begin{equation}
{1\over 2\,M_{H_Q}}<H_Q|\bar Q \;{i \over2}\, \sigma_{\mu\nu}G^{\mu\nu} Q |H_Q>
 = \mu_G^2,
\label{eq:muG}
\end{equation}
\begin{equation}
{1\over 2\,M_{H_Q}}<H_Q|{\bar Q}\,{\vec\pi}^2 Q |H_Q> = \mu_\pi^2 ,
\label{eq:mupi}
\end{equation}
The parameter $\mu_G^2$ coincides with $m_{\sigma H}^2$ introduced in
\cite{Blok2}. For mesonic states it is expressible in terms of the quantity
measured experimentally -- the hyperfine mass splittings, and it has the zero
value for baryonic states of the
type of $\Lambda_Q$; $\mu_\pi^2$ has the meaning of the average square
of spatial momentum of the heavy quark $Q$ in the hadronic state $H_Q$.
The two parameters, $\mu_G^2$ and $\mu_\pi^2$, often appear in
the combination $\mu_\pi^2-\mu_G^2$, cf. equation (\ref{eq:QQ}).

The last comment of this Section is about the operators
which are not bilinear in
$\bar{Q}$, $Q$ fields. The simplest example of appearance of such operators
is given by the diagram of Fig.2 where the heavy quark $Q$ propagates between
the current vertices. This diagram is similar to the one of Fig.1, and the
corresponding
operator follows from eq.(\ref{eq:TOPE}) by substitution
$\bar{Q}, \; Q \Rightarrow \bar{q}, \; q$, $m_q \Rightarrow m_Q$,
$k_\mu \Rightarrow q_\mu$. The additional term in $T_{\mu\nu}$ has the form
\begin{eqnarray}
\Delta T_{\mu\nu} & = & -\bar q \Gamma_\nu \frac{1}{\sla q - m_Q}
\Gamma_\mu\,q = \\
          &   & - { 2 \over (q^2-m_Q^2)}[g_{\alpha\mu}q_\nu +
g_{\alpha\nu}q_\mu - g_{\mu\nu}q_\alpha -
i \epsilon_{\mu\nu\alpha\beta}q_\beta]\,
\bar q \gamma^\alpha (1+\gamma_5)q \nonumber
\label{eq:dTOPE}
\end{eqnarray}

Matrix element of the operator $\bar{q} \gamma^\alpha q$ over $H_Q$ state
counts the number of quarks $q$ and is not small in general. The operator
coefficient
given by eq.(\ref{eq:dTOPE}) is particularly large when $q^2 \rightarrow
m_Q^2$.

In terms of the intermediate hadronic states in the forward scattering off
$H_Q$ this contribution is due to states in crossing channel containing
two $Q$ quarks - the corresponding
problem was pointed out in ref.\cite{Geor2}. The cross-channel is
not related to the weak inclusive decays under consideration and the
analysis above shows
that we can consistently omit the crossing channel together with operators
in $T_{\mu\nu}$ related to this channel. Similar consideration can be
carried out for other operators containing no heavy quark fields $Q,\;\bar{Q}$.
\section{Calculation of the differential distributions}
The differential distributions we are interested in are determined by
equation (\ref{eq:d3}) containing three invariant functions $w_1$,
$w_2$ and $w_3$. They are obtained from the results
for $h_i$ (see Appendix) by
taking the imaginary parts of the corresponding functions
(see. eq. (\ref{eq:h2w})).
The imaginary parts are due to the poles of $h_i$ and are obtained through
the relations:
\begin{equation}
{\rm Im}{1\over z^n} =
\pi\,{(-1)^{n-1} \over (n-1)!}\,{d^{n-1}\over dz^{n-1}}\,
\delta(z)
\label{eq:z2d}
\end{equation}
where $z$ is given by
\begin{equation}
z=m_Q^2-2\,m_Q\,q_0+q^2-m_c^2.
\label{eq:z}
\end{equation}
We don't present here the expression for the  triple differential
distribution which can be easily obtained by combining the equations
(\ref{eq:d3}), (\ref{eq:h2w}), (\ref{eq:z2d}) and expressions for $h_i$
from the Appendix.

Although the result is written for the physical quantity
$d^3\Gamma/ dE_e dq^2 dq_0$, it can not be directly compared with the
experimental data. An obvious signal for this is the presence of the
delta function and it's derivatives. It is not surprising because we are
sitting now right on the physical cut on mass-shell of the $q$-quark.
As we discussed in the introduction our results should be understood
in the sense of duality: that is that the predictions should be smeared
over certain duality interval.
At the moment we have no purely theoretical tools to fix the size of the
duality interval, therefore we are forced to rely on qualitative arguments
and experimental data. For example the duality interval for $q_0$ can be
inferred from the distribution in the invariant mass of the final hadronic
states. Our $\delta$-functions reflect the resonance structure at low
invariant masses. The smearing interval should be chosen in such a way
as to cover the entire resonance domain up to the onset of the smooth
behavior. Instead of smearing of the distribution one can calculate
the average characteristics like the total width $\Gamma$ or $<M_X^n>$, where
$M_X$ is the invariant mass of the final hadronic states. The power
corrections we have calculated will enter in a specific way in each particular
quantity.

Now let us proceed to the calculation of the double differential
distribution $d^2\Gamma / dE_e\,dq^2$. To this end we must
integrate over $q_0$, rather simple exercise with $\delta$-functions.
However if one would perform the integration by merely substitution
\begin{equation}
q_0\,\rightarrow q_0^*\,=\,\frac{m_Q^2+q^2-m_q^2}{2\,m_Q},
\label{eq:q*}
\end{equation}
and taking the derivatives in the case of $\delta'$ and $\delta''$,
one would get the wrong answer. The point is that integration domain
in $q_0$ has a boundary from below
\begin{equation}
q_0\,\geq E_e + \frac{q^2}{4\,E_e},
\label{eq:q0}
\end{equation}
which corresponds to $4\,E_e\,E_\nu\geq q^2$. Therefore one should
take into account the fact that $q_0$ can not cross the
boundary (\ref{eq:q0}). For that we introduce
$\theta(q_0-E_e-q^2/4E_e)$ into the integrand. The occurrence
of the $\theta$-function is important for the integration of
$\delta'(q_0-q_0^*)$ and $\delta''(q_0-q_0^*)$ which leads to appearance
of $\delta(q_0^*-E_e-q^2/4E_e)$ and $\delta'(q_0^*-E_e-q^2/4E_e)$
in the double distribution $d^2\Gamma/dq^2\,dE_e$, because of
differentiation of the $\theta$-function . The final formulae for the double
differential distribution in the lepton energy $E_e$ and $q^2$ takes
the form:
\begin{eqnarray}
\lefteqn{ \frac{d^2\Gamma}{dx dt}=                           
|V_{qQ}|^2\,\frac{G_F^2\,m_Q^5\,}{96\,\pi^3}\,x^2\,\{\, 6 \,
         (1-t)(1 - \rho - x + xt) + }                         \nonumber\\
& &   G_Q \ [\,1-5 \rho +2 t+ 10 \rho t + 10 x t - 10 x t^2 - \nonumber\\
& & \ \ (-1 + 6\rho -5 \rho^2 + x - 5\rho x + t-2 \rho t+
    5\rho^2 t + x t + 15 \rho x t + \nonumber \\
& & \ \ 5 x^2 t - 2xt^2-10 \rho x t^2- 10 x^2 t^2+5 x^2 t^3) \
      \delta((1-t)(1-x)-\rho)\,]+   \nonumber\\
& &   K_Q \ [-3 + 3 \rho + 4 t - 4 \rho t - 6 x t + 4 x t^2 - \nonumber \\
& & \ \ (1 - 2 \rho + \rho^2 - 3 x + 3 \rho x-3 t+
     2 \rho t+\rho^2 t + 11 x t - 3 \rho x t - \nonumber\\
& & \ \ 3 x^2 t - 6 x t^2 - 2 \rho x t^2 + 2 x^2 t^2 + x^2 t^3) \
        \delta((1-t)(1-x)-\rho)+ \nonumber\\
& & \ \  (1-\rho-x+xt)(1-t)(1 - 2 \rho + \nonumber\\
&&  \ \   \rho^2 - 2 x t - 2 \rho x t + x^2 t^2)
         \ \delta'((1-t)(1-x)-\rho)\,]\,\}.
\label{eq:double}
\end{eqnarray}
Here we have introduced the dimensionless variables
\begin{equation}
x=2\,E_e/m_Q , \quad t=q^2/2m_Q E_e,
\label{eq:xy}
\end{equation}
and the parameters
\begin{equation}
\rho=m_q^2/m_Q^2, \quad G_Q=\mu_G^2/m_Q^2, \quad K_Q = \mu_{\pi}^2/m_Q^2.
\label{eq:gkrho}
\end{equation}
Let us emphasize that the scale $m_Q$ used in equation (\ref{eq:xy})
is the heavy quark mass and does not coincide with $M_{H_Q}$
which is normally used in the experimental distributions.

The fact that OPE generates corrections only of the order of
${\cal O}(m_Q^{-2})$ (terms proportional to $K_Q$ and $G_Q$)
is valid for the distributions only if we use $m_Q$ as a scale,
i.e. in the variables $x,t$. Of course one can easily
rescale them to $M_{H_Q}$; then the corrections of the order of
${\cal O}(m_Q^{-1})$ will show up for trivial kinematical reasons.

We can proceed further and obtain the energy spectrum by integrating
over $q^2$. The range of integration is given by
\begin{equation}
0 \le t \le 1 - \frac{\rho}{1-x}.
\label{eq:0<y<}
\end{equation}
The result for the energy spectrum coincides with that obtained in
\cite{Bigi4}. For the sake of completeness we present it here
\footnote{Let us draw the reader's attention to the difference of notation:
$y$ in \cite{Bigi4} is equal to our $x$.}:
\begin{eqnarray}
\lefteqn{ {d\Gamma\over dx} = |V_{qQ}|^2{G_F\,m_Q^5 \over 192 \pi^3}
        \theta(1-x-\rho)2x^2\{(1-f)^2(1+2f)(2-x)+ (1-f)^3(1-x)+ } \nonumber\\
& &     (1-f)[(1-f)(2+{5\over 3}x-2f+{10\over 3} fx)-
        {f^2 \over \rho } (2x+f(12-12x+5x^2))]G_Q -               \nonumber\\
& &     [\,{5 \over 3} (1-f)^2(1+2f)x+{f^3 \over \rho} (1-f) (10x-8x^2)+
                                         \nonumber\\
& &    \,{ f^4 \over \rho^2 } (3-4f)( 2x^2-x^3 )\,]\,K_Q \},
\label{eq:dGdx}
\end{eqnarray}
where $f=\rho/(1-x)$.
Finally, performing the last integration over $x$ in the domain
\begin{equation}
0 \le x \le 1-\rho,
\label{eq:0<x<}
\end{equation}
we arrive to the total width coinciding with that in \cite{Bigi3}:
\begin{equation}
 \Gamma = |V_{qQ}|^2{G_F\,m_Q^5\over 192 \pi^3}
 [\,z_0\,(1+{1\over 2}\,(G_Q-K_Q))- 2\,z_1\,G_Q\,],
\label{eq:totalwidth}
\end{equation}
where $z_0=1-8\rho +8\rho^3-\rho^4 - 12 \rho^2 \log{\rho}$
and $z_1=(1-\rho)^4$.

Now let us discuss the characteristic features of the double distribution
(\ref{eq:double}). The most striking one is the presence of the singular
terms. The technical reason for occurrence of those terms was that we expanded
the denominator of the pole expression (\ref{eq:master}) in $\pi$
and $\sigma\,G$.
Physically this expansion reflects the shifts of the masses of particles
due to the nonperturbative effects. As it was mentioned above these
singularities reflect the structure of the resonance domain and the
predictions suitable for comparison with the experimental data require
smearing over corresponding interval. To illustrate the most salient features
of our prediction let us concentrate on the physically interesting case
of the $b\rightarrow u$ transition.

For massless $u$ quark the kinematical region of $b$ quark semileptonic
decay is shown on Fig.3. It has the form of a square with the side equal to 1
in the plane ($x=2E_e/m_b,\; t=q^2/2m_b E_e$). The right-hand
side of the square
corresponds to the maximal energy of electron $E_e=m_b/2$ while the upper
side is a maximal energy of neutrino.  In the real $B$ meson decay the
kinematical region is certainly wider; if one neglects the pion mass the
region is the square with the side $x_{max}=t_{max}=M_B/m_b$. The origin
of this window is related to the motion of the heavy $b$ quark inside of
$B$ meson. In our calculations we account for nonzero momentum of the
$b$ quark in the form of expansion which
produced singular $\delta$ and $\delta'$ terms
on the boundary. It is possible to show (see refs.\cite{Bigi4}, \cite{Jaffe})
that the expansion breaks down at distances $\sim (M_B-m_b)/m_b$ near the
boundary, so we need to average our results over the range of the order
of the window between quark and hadron boundaries. It is interesting to
note that the distribution spread off the distances of the
order $(M_B-m_b)/m_b$
while the corrections to integrals are only of the second order in $1/m_b$.

Another effect we need to account for is the structure of resonance region
near the low end of the hadronic invariant masses.
To imitate the effect let us imagine
that this region corresponds to the $u$ quark fragmentation into the hadronic
states with $s$
(the square of the invariant mass) from $s=0$ to $s=s_0=2 GeV^2$.
The curve corresponding to $s=s_0$ at Fig.3 is given by the equation:
\begin{equation}
(1-t)(1-x)= s_0/m_b^2
\label{eq:reson}
\end{equation}
and the resonance region should be included as a whole into
the process of averaging; we can predict the integral but not the structure.
\section{Application to the analysis of the experimental data}
In order to  compare our results with the
experimental data we use
the following values of parameters in eq. (\ref{eq:double}).
First, we use $m_b\sim 4.8$ GeV
as deduced from QCD analyses of the Ypsilon system and $M_B\sim 5.3$ GeV.
For the parameter $G_b $ we use the value\cite{Blok2}
\begin{equation}
G_b={3\over 4}(M^2(B^*)-M^2(B))/(4m^2_b)\sim 0.017
\label{eq:g_b}
\end {equation}
As a representative value we use for the parameter $K_b$ the value
$\sim 0.02$ (see ref. \cite{Neubert}-\cite{Ball}).

To construct the quantity convenient for the comparison of the theory
with experiment we consider the integral
\begin{equation}
P(x_c,\,t_c)={1\over \Gamma_0} \int \int_{A(x_c,t_c)} dxdt\,{d^2 \Gamma
\over dxdt}
\label{eq:plot}
\end{equation}
where $x_c,\;t_c$ is the point in ($x,\; t$) plane sitting not too close
to the boundary (outside the resonance range),
$\Gamma_0=|V_{ub}|^2 G_F^2 m_b^5/192\pi^3$ and $A(x_c,t_c)$ is the area of
integration shown on Fig.3 which includes the resonance domain.

The function $P(x, t)$ is plotted as a function of $t$ on Fig.4 for
three values of $x$ equal to 0.3, 0.6, 0.8. The last value of $x$ is close
to the border of the resonance region beyond which we cannot make reliable
predictions for the distributions considered. The dashed lines on Fig.4
describe the leading order distributions in $t$ while the solid lines
include QCD corrections we calculated. As we can see it from the curves,
the corrections are negative.

\section{Conclusion}

Let us now summarize our results. Model independent approach to
nonperturbative effects $(1/m_q)^n$ is used for calculations of
differential distributions. The effects are most pronounced near the
endpoints of the spectra. We discussed how the comparison with experiment
should be formulated accounting for the boundary effects. A somewhat
disappointing point
is that we cannot use our results to improve an extraction of $V_{ub}$ by
the consideration of $q^2$  dependence. Indeed, experimentally the signal of
$b \rightarrow u$ is
due to the range of electron energy $E_e$ near the upper end where
$b \rightarrow c$ is absent. However as it follows from Fig.3 the
distribution in $q^2$ at such energies is concentrated in the resonance
domain, and no model-independent prediction emerges.
\newpage
\noindent {\bf Acknowledgments}

The authors are grateful to Dr.~R.~Poling for stimulating discussions
of experimentally measured distributions. Those discussions initiated
the present study. We also benefitted from helpful discussions with
Dr.~N.~Uraltsev and Dr.~M.~Voloshin. This work was supported in part
by DOE under the grant number DOE-AC02-83ER40105.

After our paper was completed, one of us -- B.B. -- learned
from A. Manohar about the similar work,
to be published \cite{Manohar}.

\newpage
\appendix
\renewcommand{\theequation}{A.\arabic{equation}}
\setcounter{equation}{0}
\section{Hadronic invariant functions}
\par Here we present the results of calculations of different hadronic
invariant functions $h_i$, introduced by eqs.(\ref{eq:h2h}, (\ref{eq:full})).
The structure functions $w_i$ are simply related to $h_i$ by eqs.(\ref{eq:h2w})
and (\ref{eq:z2d}).
We use the following notation : $q_0=q \cdot v$, $\vec q\,^2 = q_0^2-q^2$
and $z = m_Q^2-2\,m_Q\,q_0 + q^2 -m_q^2$.

For the Vector$\times$Vector functions we have:
\begin{eqnarray}
\lefteqn{ h_1^{VV}=
    -\,[\,(\,m_Q -m_q - q_0\,) - (\mu_G^2 - \mu_{\pi}^2\,){1\over 2 m_Q}\,
       ({1\over 3}+{m_q\over m_Q}\,)\,]\,{1\over z} - } \nonumber \\
& & \;\;\; {1\over m_Q}\,[\,{1\over 3}\,\mu_G^2\,
      (\,(4\,m_Q - 3 q_0)(\,m_Q - m_q - q_0) + 2\,\vec q\,^2\,)+ \nonumber \\
& & \;\;\;\;\;\;\; \mu_{\pi}^2\,(q_0\,(m_Q -m_q-q_0) -
                  {2\over 3}\,\vec q\,^2\,)\,]\,{1\over z^2}\,-  \nonumber \\
& & \;\;\;\;\;\;\;\;\; {4\over 3}\,\mu_{\pi}^2\,\vec q\,^2\,
       (m_Q - m_q - q_0)\,{1\over z^3}\,\,,
\label{eq:vv1}
\end{eqnarray}
\begin{eqnarray}
\lefteqn{ h_2^{VV} = -\,[\,2\,m_Q -
       {5\over 3m_Q}\,(\mu_G^2-\mu_{\pi}^2)\,]\,\,{1\over z} - } \nonumber\\
& & \;\;\;\;\;\;\;\;\;
  {2\over 3}\,[\,2 \mu_G^2\,(m_Q-m_q) -5\,\mu_G^2\,q_0 +
                         7\,\mu_{\pi}^2\,q_0 ]\,{1\over z^2} \,- \nonumber\\
& & \;\;\;\;\;\;\;\;\;{8\over 3} \,m_Q\,\mu_{\pi}^2\,\vec q\,^2\,{1\over z^3}
\label{eq:vv2}
\end{eqnarray}
\begin{equation}
 h_3^{VV} = 0\,,
 \label{eq:vv3}
\end{equation}
\begin{equation}
h_4^{VV}= - {4\over 3 m_Q}\,(\mu_{\pi}^2 - \mu_G^2)\,{1\over z^2}\,,
\label{eq:vv4}
\end{equation}
\begin{equation}
h_5^{VV} =  {1\over z} -
        {1\over 3}\,[\,5\,{q_0\over m_Q}(\mu_G^2\, - \mu_{\pi}^2\,) -
        4\,\mu_{\pi}^2)\,]\,{1\over z^2} +
 {4\over 3}\,\mu_{\pi}^2\,\vec q\,^2\,{1\over z^3}\,.
\label{eq:vv5}
\end{equation}
\indent To get the functions $h_i^{AA}$ for Axial$\times$Axial tensor from
$h_i^{VV}$ one
should substitute $m_q$ by $(-\,m_q)$ in eqs.(\ref{eq:vv1} - \ref{eq:vv5}).

For the Axial$\times$Vector tensor only
one invariant structure survives:
\begin{equation}
h^{VA}_3=
         \,{1\over z}+
         [\,2\,\mu_G^2 + {5\over 3}\,(\mu_{\pi}^2 -
        \mu_G^2)\,{q_0\over m_Q} \,]\,{1\over z^2}\,+
        {4\over 3}\,\mu_{\pi}^2\,\vec q\,^2\,{1\over z^3}\,.
\label{eq:va1}
\end{equation}

\indent Summing up we get
the result for the full hadronic tensor $h_{\mu\nu}$.
\begin{eqnarray}
\lefteqn{ h_1 = - \,[\,2\,(m_Q - q_0) -
   {1\over 3 m_Q}\,(\mu_G^2 - \mu_{\pi}^2)\,]\,{1\over z} - } \nonumber\\
& &\;\;\;\; [\,{2\over 3 m_Q}\,\mu_G^2\,(4\,m_Q^2 +
            2\,\vec q\,^2 - 7\,m_Q\,q_0 + 3\,q_0^2) +  \nonumber\\
& &\;\;\;\; {\mu_{\pi}^2 \over 2 m_Q}\,(4\,q_0\,(m_Q - q_0) -
                 {8\over3}\,\vec q\,^2 \,)\,]\,{1\over z^2} - \nonumber\\
& &\;\;\;\; {8\over3}\,\mu_{\pi}^2\,\vec q\,^2\,
              (m_Q - q_0)\,{1\over z^3}\,,
\label{eq:hh1}
\end{eqnarray}
\begin{eqnarray}
\lefteqn{h_2 = - \,
      [\,4\,m_Q + {10\over 3m_Q}\,(\mu_{\pi}^2 - \mu_G^2)\,]\,{1\over z} - }
      \nonumber\\
& &\;\;\;\; [{28\over 3}\,\mu_{\pi}^2\,q_0 + \mu_G^2\,({8\over 3}\,m_Q -
      {20\over 3}\,q_0)\,]\,{1\over z^2}\,- \nonumber\\
& &\;\;\;\; {16\over 3}\,\mu_{\pi}^2\,m_Q\,\vec q\,^2\,{1\over z^3}\,.
\label{eq:hh2}
\end{eqnarray}
\begin{equation}
h_3 = - \,2\,{1\over z} -
      [\,4\,\mu_G^2 + {10\over3}\,(\mu_{\pi}^2 -
       \mu_G^2\,)\,{q_0\over m_Q}\,]\,{1\over z^2} -
     {8\over3}\,\mu_{\pi}^2\,\vec q\,^2\,{1\over z^3}\,,
\label{eq:hh3}
\end{equation}
\begin{equation}
h_4 =
  - {8\over 3 m_Q}\,(\mu_{\pi}^2 - \mu_G^2)\,{1\over z^2}\,,
\label{eq:hh4}
\end{equation}
\begin{equation}
h_5 = 2\,{1\over z} -
     {2\over3}\,[\,5\,(\mu_G^2- \mu_{\pi}^2)\,{q_0\over m_Q} -
     4\,\mu_{\pi}^2)\,]\,{1\over z^2} +
     {8\over3}\,\mu_{\pi}^2\,\vec q\,^2\,{1\over z^3}\,,
\label{eq:hh5}
\end{equation}

\vspace{0.5in}
\noindent {\Large \bf Figure captions}

\indent {\bf Fig.1:} The tree diagram determining the transition operator
$T_{\mu\nu}$
in the leading approximation. The dashed lines correspond to the weak currents,
the solid internal line describes the propagation of the quark $q$ and the
bold external lines represent the heavy quark $Q$.

{\bf Fig.2:} The tree diagram determining the operator without the heavy quark
$Q$.  Now the bold internal line describes the propagation of the heavy quark
$Q$ and the  solid external lines represent the quark $q$.

{\bf Fig.3:} The kinematical region of the decay for $b \rightarrow u$ decays
in coordinates $x=2E_e/m_b$ and $t=q^2/2m_b E_e$. The solid lines are
the kinematical boundary for the $b$ quark decay ($x_{max}=t_{max}=1$) and
the dashed lines are the boundary for $B$ meson decay
($x_{max}=t_{max}=M_B/m_b$). The area of integration for the distribution
$P(x,t)$ (see eq.(\ref{eq:plot})) is shadowed. It includes integration over
the resonance domain.

{\bf Fig.4:} The integrated distribution $P(x,t)$ (see eq.(\ref{eq:plot}))
for the case $b \rightarrow u$ is plotted as function of $t=q^2/2m_b E_e$
for few values of $x=2E_e/m_b$. The dashed lines correspond to the
leading order distribution while the solid lines account for QCD corrections.
The lines stop at the border of resonance region.
The corrections are negative and their relative magnitude becomes larger
when $x$ approaches the resonance region.


\newpage
\begin{thebibliography}{99}
%
\bibitem{Alta1}
G. Altarelli, N. Cabibbo, G. Corbo, L. Maiani and G. Martinelli, {\em Nucl.
Phys.}, {\bf B208} (1982) 365.

\bibitem{Isgu1}
N. Isgur, D. Scora, B. Grinstein and M. Wise,
{\em Phys. Rev.} {\bf D39} (1989) 799.

\bibitem{Volo1}
M. Voloshin and M. Shifman, {\em Yad. Fiz.} {\bf 41} (1985) 187
[{\em Sov. J. Nucl. Phys.} {\bf 41} (1985) 120];
{\em ZhETF} {\bf 91} (1986) 1180
[{\em Sov. Phys. JETP} {\bf 64} (1986) 698].

\bibitem{Wils1}
K. Wilson, {\em Phys. Rev.} {\bf 179} (1969) 1499; \\
K. Wilson and J. Kogut, {\em Phys. Rep.} {\bf 12} (1974) 75.



\bibitem{Bigi1}
I. Bigi, N. Uraltsev and A. Vainshtein,
{\em Phys. Lett.} {\bf B293} (1992) 430.

\bibitem{Blok1}
B. Blok and M. Shifman, {\em Nucl. Phys.} {\bf B399} (1993) 441; 459.

\bibitem{Bigi4}
I. Bigi, M. Shifman, N. Uraltsev and A. Vainshtein,
{\em "QCD Predictions For Lepton Spectra in Inclusive Heavy
Flavor Decays "}, March 1992, Preprint TPI-MINN-93/12-T [Phys. Rev. Lett.,
submitted]. More detailed text is in preparation.

\bibitem{Bigi3}
I. Bigi, B. Blok, M. Shifman, N. Uraltsev and A. Vainshtein,
{\em "A QCD 'Manifesto' on Inclusive Decays of Beauty and Charm"},
Talk at DPF Meeting of APS, November 1992, Preprint TPI-MINN-92/67-T.

\bibitem{Geor2}
J. Chay, H. Georgi and  B. Grinstein, {\em  Phys. Lett.} {\bf B247} (1990)
399.

\bibitem{Shif1}
M. Shifman, A. Vainshtein and V. Zakharov,
{\em Nucl. Phys.} {\bf B147} (1979) 385.

\bibitem{Bjor1}
J. D. Bjorken, Invited Talk at {\em Les Rencontres de la Valle d'Aosta,
La Thuille, 1990}, Preprint SLAC-PUB-5278, 1990.

\bibitem{Geor1}
E. Eichten and B. Hill, {\em Phys. Lett.} {\bf B234} (1990) 511;\\
H. Georgi, {\em Phys. Lett.} {\bf B240} (1990) 447.

\bibitem{Novi1}
V. Novikov, M. Shifman, A. Vainshtein and V. Zakharov,
{\em Fortsch. Phys.} {\bf B32} (1984) 585.

\bibitem{Blok2}
B. Blok and M. Shifman,
{\em Nucl. Phys.} {\bf B389} (1993) 534.


\bibitem{Neubert}
M. Neubert, preprint SLAC-PUB-5770 (1992)

\bibitem{Bagan}
E. Bagan et al., {\em Phys. Lett.} {\bf B278} (1992) 467

\bibitem{Ball}
P. Ball et al., Preprint TUM-T31-31-92

\bibitem{Jaffe}
R.L.Jaffe and L.Randall,
{\em "Heavy Quark Fragmentation Into Heavy Mesons"},
Preprint CTP \# 2189, May 1993

\bibitem{Manohar}
A. Manohar and M. Wise, In preparation

\end{thebibliography}
\end{document}